\documentclass[twocolumn,showpacs,preprintnumbers,amsmath,amssymb,prb]{revtex4-1}

\usepackage{graphicx}% Include figure files
\usepackage{dcolumn}% Align table columns on decimal point
\usepackage{bm}% bold math
\usepackage{tabularx}% Table width
\usepackage{amsmath}% Higher math

\begin{document}

\title{Multiband nodeless superconductivity near the charge-density-wave quantum critical point in ZrTe$_{3-x}$Se$_x$}

\author{Shan Cui,$^1$ Lan-Po He,$^1$ Xiao-Chen Hong,$^1$ Xiang-De Zhu,$^{2,4}$ Cedomir Petrovic,$^4$ and Shi-Yan Li$^{1,3,*}$}

\affiliation{$^1$ State Key Laboratory of Surface Physics, Department of Physics, and Laboratory of Advanced Materials, Fudan University, Shanghai 200433, China\\
$^2$ High Magnetic Field Laboratory, Chinese Academy of Sciences and University of Science and Technology of China, Hefei 230031, China\\
$^3$Collaborative Innovation Center of Advanced Microstructures, Fudan University, Shanghai 200433, China\\
$^4$Condensed Matter Physics and Materials Science Department, Brookhaven National Laboratory, Upton, New York 11973, USA}

\date{\today}

\begin{abstract}
  Recently it was found that selenium doping can suppress the charge-density-wave (CDW) order and induce bulk superconductivity in ZrTe$_3$. The observed superconducting dome suggests the existence of a CDW quantum critical point (QCP) in ZrTe$_{3-x}$Se$_x$ near $x \approx$ 0.04. To elucidate its superconducting state near the CDW QCP, we measure the thermal conductivity of two ZrTe$_{3-x}$Se$_x$ single crystals ($x$ = 0.044 and 0.051) down to 80 mK. For both samples, the residual linear term $\kappa_0/T$ at zero field is negligible, which is a clear evidence for nodeless superconducting gap. Furthermore, the field dependence of $\kappa_0/T$ manifests multigap behavior. These results demonstrate multiple nodeless superconducting gaps in ZrTe$_{3-x}$Se$_x$, which indicates conventional superconductivity despite of the existence of a CDW QCP.
\end{abstract}


\maketitle

\section{Introduction}

Charge-density-wave (CDW) order usually exists in some low-dimensional materials, especially those transition-metal chalcogenides. \cite{Wilson,Kim,Salvo,Boswell} When the CDW order is suppressed by doping or pressure, a list of them can be tuned to superconductors. \cite{Morosan,Kusmartseva,Sipos,Hoesch} In the temperature-doping ($T$-$x$) or temperature-pressure ($T$-$p$) phase diagram, sometimes a superconducting dome is observed on top of a CDW quantum critical point (QCP). \cite{Morosan,Kusmartseva,Sipos,Hoesch} The reminiscent of this kind of phase diagram to the heavy-fermion and high-$T_c$ cuprate superconductors raises the possibility of unconventional superconductivity caused by CDW fluctuations. \cite{Morosan,Kusmartseva,Sipos,Hoesch,Norman}

ZrTe$_3$ is such a compound in which CDW order and superconductivity compete and coexist. \cite{Yamaya1} It belongs to a family of trichalcogenides MX$_3$ (M = Ti, Zr, Hf, U, Th, and X = S, Se, Te). The structure consists of infinite X-X chains formed by stacking MX$_3$ prisms. \cite{Furuseth} The polyhedra are arranged in double sheets and stacked along monoclinic $c$-axis by van der Waals forces. \cite{Furuseth} Pristine ZrTe$_3$ itself harbors filamentary superconductivity with $T_c \sim$ 2 K.\cite{Yamaya1} The CDW vector $\overrightarrow{q} \approx (1/14; 0; 1/3)$ is developed in ZrTe$_3$ below $T_{CDW} \sim$ 63 K.\cite{Eaglesham} Like other CDW materials, pressure and doping can melt the CDW order and stabilize its superconductivity to bulk. \cite{Zhu1,Lei,Yamaya2,Zhu2} Recently, isovalent substitution of Se for Te is also found to cause a superconducting dome in ZrTe$_{3-x}$Se$_x$ system, with maximum $T_c$ = 4.4 K at the optimal doping $x$ = 0.04. \cite{Zhu3} It was suggested that this superconductivity may be mediated by quantum critical charge fluctuations. \cite{Zhu3} To clarifying the underlying pairing mechanism, it is important to know the superconducting gap symmetry and structure.

Ultra-low-temperature heat transport is an established bulk technique to probe the superconducting gap structure \cite{Shakeripour}. The existence of a finite residual linear term $\kappa_0/T$ in zero magnetic field is an evidence for gap nodes \cite{Shakeripour}. The field dependence of $\kappa_0/T$ may further give support for a nodal superconducting state, and provide information on the gap anisotropy, or multiple gaps \cite{Shakeripour}.

In this paper, we measure the ultra-low-temperature thermal conductivity of ZrTe$_{3-x}$Se$_x$ single crystals near optimal doping, to investigate whether its superconducting state is unconventional. The negligible $\kappa_{0}/T$ in zero field and the rapid field dependence of $\kappa_{0}(H)/T$ in low field strongly suggest multiple nodeless superconducting gaps in ZrTe$_{3-x}$Se$_x$. In this sense, the superconductivity in ZrTe$_{3-x}$Se$_x$ is likely conventional.

\section{Experiment}

The ZrTe$_{3-x}$Se$_x$ single crystals were grown by iodine vapor transport method. \cite{Zhu1,Zhu3} Two single crystals from different batches, both with nominal composition $x$ = 0.04, were used for this study. Their exact compositions were determined by wavelength-dispersive spectroscopy (WDS), utilizing an electron probe microanalyzer (Shimadzu EPMA-1720). The dc magnetization was measured at $H$ = 20 Oe, with zero-field cooling, using a SQUID (MPMS, Quantum Design). The samples were cleaved and cut to rectangular bars, with typical dimensions of 2.12 $\times$ 1.01 $\times$ 0.030 mm$^3$. The largest surface is $ab$-plane. Contacts were made directly on the sample surfaces with silver paint, which were used for both resistivity and thermal conductivity measurements. The contacts are metallic with typical resistance 200 m$\Omega$ at 2 K. In-plane thermal conductivity was measured in a dilution refrigerator, using a standard four-wire steady-state method with two RuO$_2$ chip thermometers, calibrated {\it in situ} against a reference RuO$_2$ thermometer. Magnetic fields were applied along the $c$ axis and perpendicular to the heat current. To ensure a homogeneous field distribution in the sample, all fields were applied at temperature above $T_c$.

\section{Results and discussion}

According to the WDS results, the actual Se content of the two ZrTe$_{3-x}$Se$_x$ single crystals is $x$ = 0.044 and 0.051, respectively. Below we will use the actual $x$. Figure 1(a) presents the normalized dc magnetization of ZrTe$_{2.956}$Se$_{0.044}$ and ZrTe$_{2.949}$Se$_{0.051}$ single crystals. The $T_c$ defined by the onset of diamagnetic transition is 4.0 K for both samples. The significant diamagnetic response confirms that the superconductivity is stabilized to bulk from the filament superconductivity in pristine ZrTe$_3$, which is consistent with previous report. \cite{Zhu3} This bulk superconductivity will be further supported by our thermal conductivity data in this study.

\begin{figure}
  \includegraphics[clip,width=7.8cm]{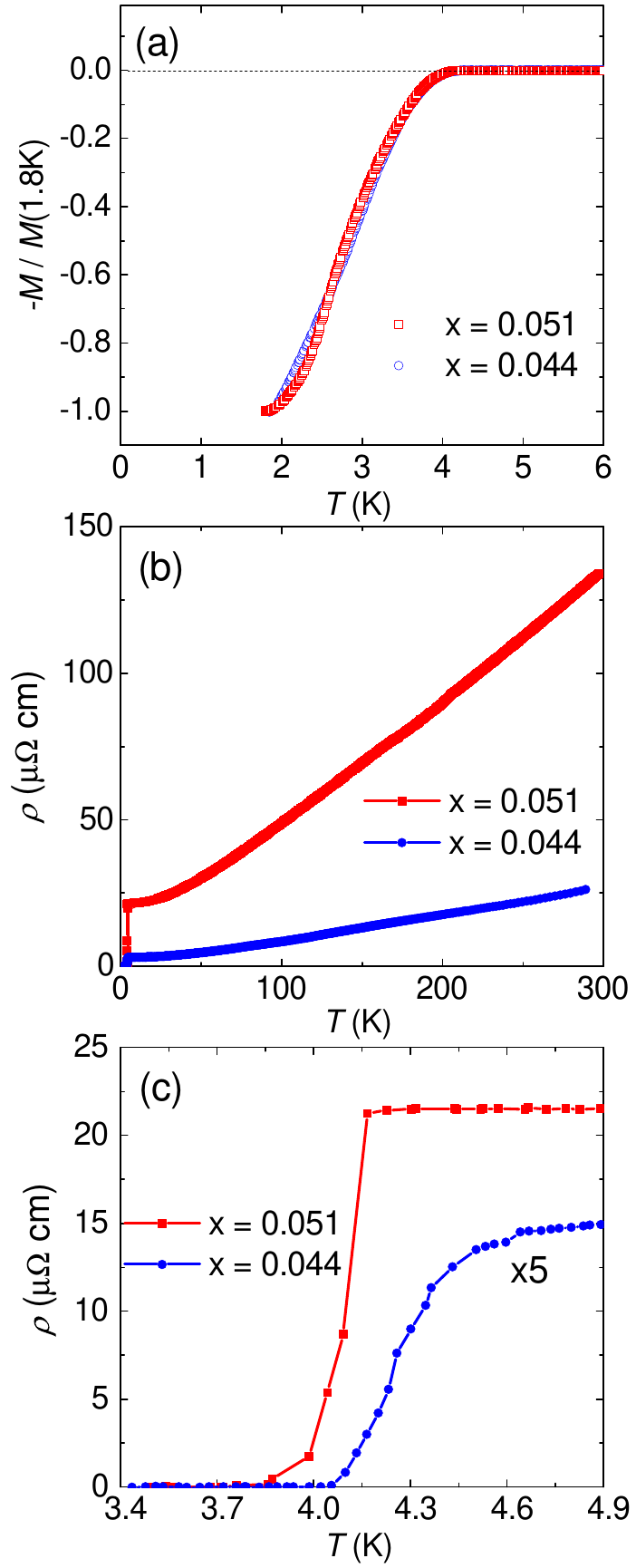}
  \caption{(Color online) (a) The normalized dc magnetization of ZrTe$_{2.956}$Se$_{0.044}$ and ZrTe$_{2.949}$Se$_{0.051}$ single crystals, measured in $H$ = 20 Oe with zero-field-cooled (ZFC) process. (b) The in-plane resistivity of ZrTe$_{2.956}$Se$_{0.044}$ and ZrTe$_{2.949}$Se$_{0.051}$ single crystals. No anomaly was observed in the normal state, suggesting the complete suppression of CDW state. (c) The resistive superconducting transition at low temperature. For clarity, the resistivity value of the $x$ = 0.044 sample is magnified by five times. The $T_c$ defined by $\rho$ = 0 is 4.06 and 3.87 K for $x$ = 0.044 and 0.051 samples, respectively.}
\end{figure}

Figure 1(b) shows the in-plane resistivity $\rho(T)$ of ZrTe$_{2.956}$Se$_{0.044}$ and ZrTe$_{2.949}$Se$_{0.051}$ single crystals. No anomaly was observed in the normal state, suggesting the complete suppression of CDW state in them. \cite{Zhu3} Fitting the normal-state resistivity data below 60 K to $\rho(T)$ = $\rho_0$ + $AT^n$ gives residual resistivity $\rho_0$ = 2.82 and 21.5 $\mu\Omega$ cm for $x$ = 0.044 and 0.051 samples, respectively. The resistive superconducting transition at low temperature is plotted in Fig. 1(c). The $T_c$ defined by $\rho$ = 0 is 4.06 and 3.87 K for $x$ = 0.044 and 0.051 samples, respectively. Both of them are near the optimal doping in the phase diagram of ZrTe$_{3-x}$Se$_x$, and the $x$ = 0.051 sample is slightly overdoped. \cite{Zhu3}

To determine their upper critical field $H_{c2}$, the low-temperature resistivity of these two samples under magnetic fields was also measured. Figure 2(a) and 2(b) show the low temperature $\rho(T)$ curve of ZrTe$_{2.956}$Se$_{0.044}$ and ZrTe$_{2.949}$Se$_{0.051}$ single crystals under various fields. With increasing field, the superconducting transition is gradually suppressed to lower temperature, and the magnetoresistance in the normal state is very weak. The $H_{c2}(T)$, defined by $\rho = 0$ in (a) and (b), is plotted in Fig. 2(c) for both $x$ = 0.044 and 0.051 samples. From Fig. 2(c), we roughly estimate $H_{c2}(0) \approx$ 1.40 and 0.85 T for them, respectively.

\begin{figure}
  \includegraphics[clip,width=7cm]{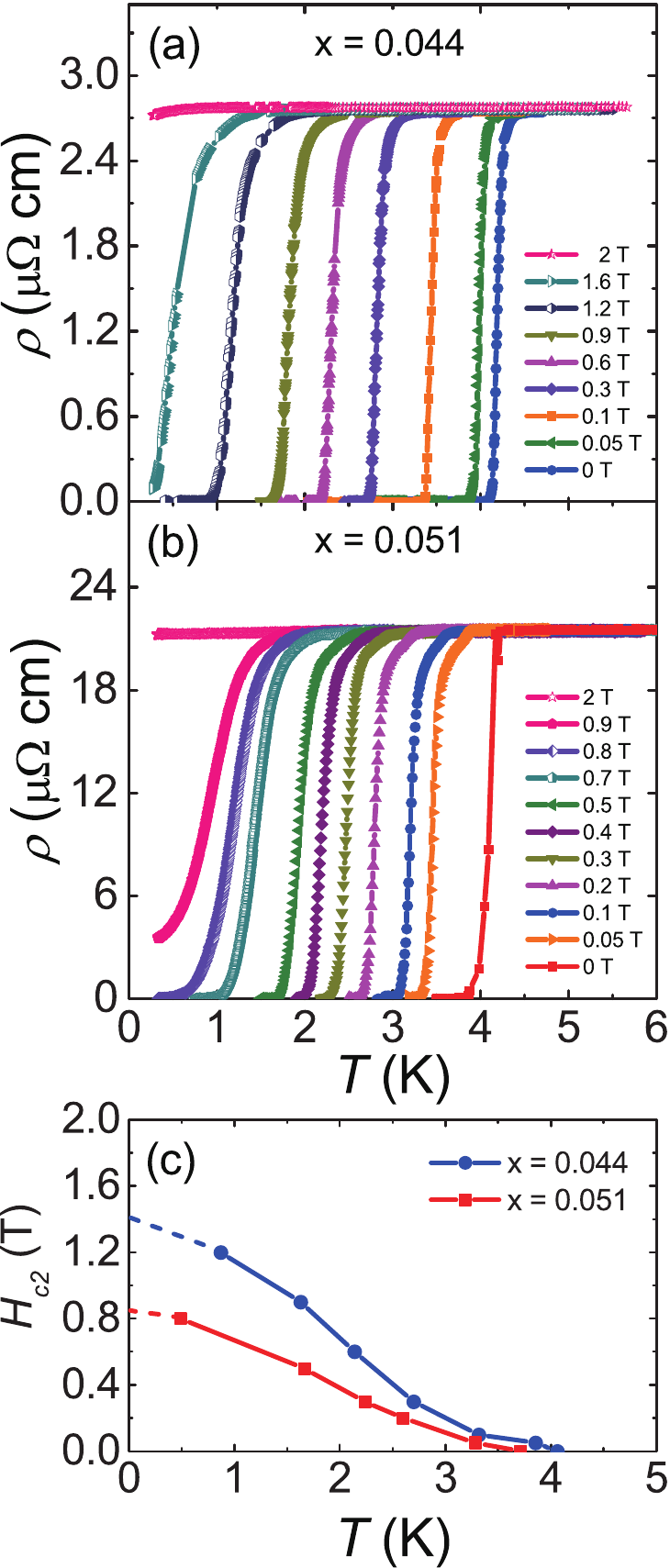}
  \caption{(Color online) Low-temperature resistivity of (a) ZrTe$_{2.956}$Se$_{0.044}$ and (b) ZrTe$_{2.949}$Se$_{0.051}$ single crystals under various magnetic fields. (c) Temperature dependence of the upper critical field $H_{c2}(T)$, defined by $\rho = 0$ in (a) and (b). The dashed lines are guide to eye, which point to $H_{c2}(0) \approx$ 1.40 and 0.85 T for $x$ = 0.044 and 0.051 samples, respectively.}
\end{figure}

The temperature dependence of in-plane thermal conductivity for ZrTe$_{2.949}$Se$_{0.044}$ and ZrTe$_{2.956}$Se$_{0.051}$ single crystals in zero and applied magnetic fields is shown in Fig. 3, plotted as $\kappa/T$ vs $T$. The thermal conductivity at very low temperature can usually be fitted to $\kappa/T$ = $a + bT^{\alpha-1}$. \cite{Sutherland,SYLi} The two terms $aT$ and $bT^{\alpha}$ represent contributions from electrons and phonons, respectively. The power $\alpha$ is typically between 2 and 3, due to specular reflections of phonons at the boundary. \cite{Sutherland,SYLi} One can see that all the curves in Fig. 3 are roughtly linear, therefore we fix $\alpha$ to 2. In zero field, the fittings give $\kappa_0/T$ = 0.008 $\pm$ 0.008 and 0.009 $\pm$ 0.002 mW K$^{-2}$ cm$^{-1}$ for the $x$ = 0.044 and 0.051 samples, respectively. Such a tiny $\kappa_0/T$ in zero field is negligible for both samples. As $T \to 0$, since all electrons become Cooper pairs for $s$-wave nodeless superconductors, there are no fermionic quasiparticles to conduct heat. Therefore there is no residual linear term of $\kappa_0/T$, as seen in V$_3$Si \cite{Sutherland}. However, for unconventional superconductors with nodes in the superconducting gap, the nodal quasiparticles will contribute a finite $\kappa_0/T$ in zero field. \cite{Shakeripour} For example, $\kappa_0/T$ = 1.41 mW K$^{-2}$ cm$^{-1}$ for the overdoped cuprate Tl$_{2}$Ba$_{2}$CuO$_{6+\delta}$ (Tl-2201), a $d$-wave superconductor with $T_c$ = 15 K. \cite{Proust} For the $p$-wave superconductor Sr$_2$RuO$_4$, $\kappa_0/T$ = 17 mW K$^{-2}$ cm$^{-1}$. \cite{Suzuki} Therefore, the negligible $\kappa_0/T$ of the $x$ = 0.044 and 0.051 samples samples suggest that the superconducting gap of ZrTe$_{3-x}$Se$_x$ is nodeless. Note that the negligible $\kappa_0/T$ in zero field also supports the bulk superconductivity in our samples.

\begin{figure}
  \includegraphics[clip,width=7.8cm]{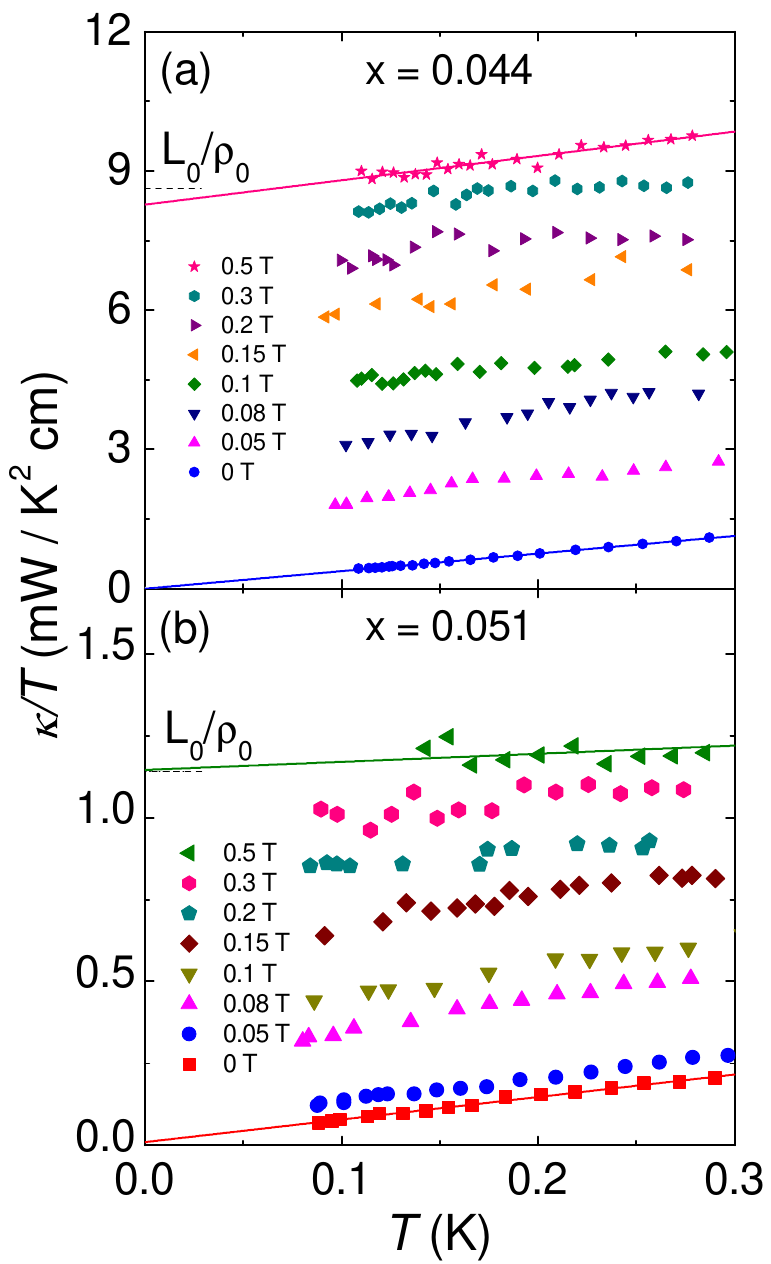}
  \caption{(Color online) Low-temperature in-plane thermal conductivity of (a) ZrTe$_{2.956}$Se$_{0.044}$ and (b) ZrTe$_{2.949}$Se$_{0.051}$ single crystals in zero and magnetic fields. The lines are fits of the data to $\kappa/T = a + bT^{\alpha-1}$, with $\alpha$ fixed to 2. The dashed lines represent the normal-state Wiedemann-Franz law expectation $L_0$/$\rho_0$ for the $x$ = 0.044 and 0.051 samples, respectively.}
\end{figure}

When applying field, $\kappa/T$ gradually increases with increasing field, as seen in Fig. 3. In $H$ = 0.5 T, the fittings give $\kappa_0/T$ = 8.27 $\pm$ 0.08 and 1.14 $\pm$ 0.03 mW K$^{-2}$ cm$^{-1}$ for the $x$ = 0.044 and 0.051 samples, respectively. These values roughly meet their Wiedemann-Franz law expectations $L_0/\rho_0$ ($L_0$ is the Lorenz number 2.45 $\times$ 10$^{-8}$ W $\Omega$ K$^{-2}$ and $\rho_0$ is the sample's residual resistivity). The verification of the Wiedemann-Franz law in the normal state shows the reliability of our thermal conductivity measurements. The bulk $H_{c2}$(0) $\approx$ 0.5 T is taken for both samples, which is lower than those determined from resistivity measurements.

To gain more information of the gap structure in ZrTe$_{2.956}$Se$_{0.044}$ and ZrTe$_{2.949}$Se$_{0.051}$, we check the field dependence of their $\kappa_0/T$. The normalized $\kappa_0/T$ as a function of $H/H_{c2}$ is plotted in Fig. 4. For comparison, the data of the clean $s$-wave superconductor Nb, \cite{Lowell} the multiband $s$-wave superconductor NbSe$_2$, \cite{Boaknin} and an overdoped sample of the $d$-wave superconductor Tl-2201 are also plotted. \cite{Proust} The slow field dependence of $\kappa_0/T$ at low field for Nb manifests its single isotropic superconducting gap. From Fig. 4, the curves of $x$ = 0.044 and 0.051 samples are similar to that of NbSe$_2$, a multiband $s$-wave superconductor with the gap ratio $\Delta_l/\Delta_s$ $\approx$ 3. \cite{Boaknin} This suggests that ZrTe$_{3-x}$Se$_x$ also has multiple nodeless superconducting gaps. Previously, {\it ab initio} calculation of the band structure for ZrTe$_3$ at ambient pressure gives a central rounded 2D Fermi surface sheet and two flatter q1D sheets. \cite{Hoesch} Therefore, the observation of multiple nodeless superconducting gaps in ZrTe$_{3-x}$Se$_x$ system is not surprising.

\begin{figure}
  \includegraphics[clip,width=7cm]{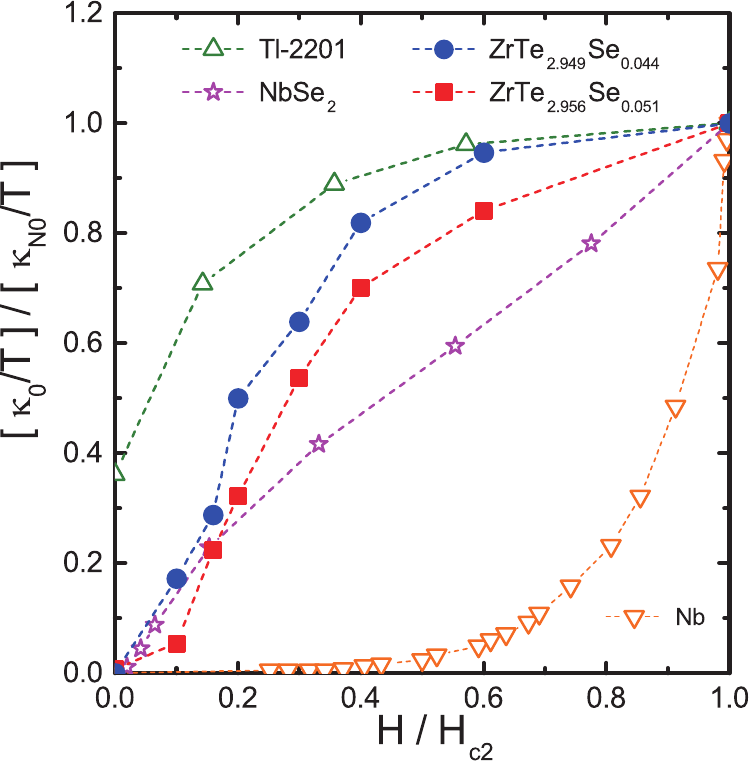}
  \caption{(Color online) Normalized residual linear term $\kappa_0/T$ of ZrTe$_{2.956}$Se$_{0.044}$ and ZrTe$_{2.949}$Se$_{0.051}$ single crystals as a function of $H/H_{c2}$. Similar data of the clean $s$-wave superconductor Nb, \cite{Lowell} an overdoped $d$-wave cuprate superconductor Tl-2201, \cite{Proust} and the multiband $s$-wave superconductor NbSe$_2$ \cite{Boaknin} are also plotted for comparison.}
\end{figure}

Theoretically, it has been shown that unconventional superconductivity with $d_{xy}$ symmetry can appear in close proximity to a charge-ordered phase, and the superconductivity is mediated by charge fluctuations. \cite{Scalapino,Merino} Since the $d_{xy}$-wave gap has line nodes, our results clear rule out this kind of unconventional superconductivity in ZrTe$_{3-x}$Se$_x$. In this context, the superconductivity in ZrTe$_{3-x}$Se$_x$ is likely conventional. Similar situation happens in Cu$_x$TiSe$_2$ system. Thermal conductivity measurements suggested conventional $s$-wave superconductivity with single isotropic gap in Cu$_{0.06}$TiSe$_2$, near where the CDW order vanishes. \cite{Li} So far, the evidence for unconventional superconductivity induced by CDW fluctuations in real materials is still lack. The experiments on more systems with superconductivity near a CDW QCP are needed.

\section{Conclusion}
In summary, we have measured the ultra-low-temperature thermal conductivity of ZrTe$_{2.956}$Se$_{0.044}$ and ZrTe$_{2.949}$Se$_{0.051}$ single crystals, which are near the optimal doping in the phase diagram of ZrTe$_{3-x}$Se$_x$ system. The absence of $\kappa_0/T$ in zero field for both compounds gives strong evidence for nodeless superconducting gap. The field dependence of $\kappa_0(H)/T$ further suggests multiple nodeless gaps in ZrTe$_{3-x}$Se$_x$. Unconventional superconductivity with line nodes is excluded in this trichalcogenide system although there is a CDW QCP. It is likely that the superconductivity in ZrTe$_{3-x}$Se$_x$ is still conventional.

\begin{center}
{\bf ACKNOWLEDGEMENTS}
\end{center}

This work is supported by the Ministry of Science and Technology of China (National Basic Research Program No. 2012CB821402 and No. 2015CB921401), the Natural Science Foundation of China (No. 91421101, No. 11422429, and No. 11204312), the Program for Professor of Special Appointment (Eastern Scholar) at Shanghai Institutions of Higher Learning, and STCSM of China (No. 15XD1500200).\\

$^*$ E-mail: shiyan$\_$li@fudan.edu.cn

\end{document}